\documentclass[amsart,twocolumn,prl,secnumarabic,amssymb,nobibnotes,amsmath,aps,showpacs]{revtex4-1}

\usepackage{graphics}
\usepackage{graphicx}
\usepackage{amssymb}
\usepackage{braket}

\usepackage[usenames,dvipsnames]{color}

\begin{document}

\bibliographystyle{apsrev4-1}

\title{Bright, continuous beams of cold free radicals}

\author{J.~C. Shaw}
\affiliation{Department of Physics, University of Connecticut, 196 Auditorium Road, Unit 3046, Storrs, Connecticut 06269, USA.}

\author{D.~J. McCarron}
\email{daniel.mccarron@uconn.edu}
\affiliation{Department of Physics, University of Connecticut, 196 Auditorium Road, Unit 3046, Storrs, Connecticut 06269, USA.}

\begin{abstract}
We demonstrate a cryogenic buffer gas-cooled molecular beam source capable of producing bright, continuous beams of cold and slow free radicals via laser ablation over durations of up to 60~seconds. The source design uses a closed liquid helium reservoir as a large thermal mass to minimize heating and ensure reproducible beam properties during operation. Under typical conditions, the source produces beams of our test species SrF, containing $5\times10^{12}$~molecules per steradian per second in the X$^{2}\Sigma(v=0, N=1)$ state with a rotational temperature of 1.0(2)~K and a forward velocity of $140~$m/s. The beam properties are robust and unchanged for multiple cell geometries but depend critically on the helium buffer gas flow rate, which must be $\geq10$~standard cubic centimeters per minute to produce bright, continuous beams of molecules for an ablation repetition rate of 55~Hz.
\end{abstract}

$\pacs{}$

\maketitle

% Introduction
Beams of cold and slow molecules from cryogenic buffer gas sources have played a central role in recent improved precision measurements \cite{Baron2014,Andreev2018}, high-resolution spectroscopy \cite{Norrgard2017,Truppe2019} and the direct laser cooling and trapping of molecules at ultracold temperatures \cite{McCarron2018,Tarbutt2018}. Direct cooling methods for molecules have the potential to produce a chemically diverse range of diatomic and polyatomic species at ultracold temperatures which are well-suited for proposed applications including tests of fundamental physics \cite{Kozyryev2017b}, and controlled chemistry \cite{Wells2011b}.

Cryogenic buffer gas beam sources rely on a flow of cold inert gas, usually helium or neon, to sympathetically cool the molecular species of interest and collapse the occupied rovibrational state distribution \cite{Maxwell2005,Barry2011,Hutzler2011,Hutzler2011b}. This thermalization takes place inside an enclosed cell as the species of interest becomes entrained within the inert gas flow and exits the cell through a small hole to form a beam. These molecular beams have forward velocities between $\sim50~$ and $200~$m/s and advances in slowing techniques using radiation pressure \cite{Barry2012,Truppe2017b} have enabled molecules below $\sim10~$m/s to be captured and cooled by magneto-optical traps (MOTs) \cite{Barry2014}. Today's molecular MOTs provide confining and damping forces comparable to those in atomic MOTs but can only capture $10^{4}-10^{5}$ molecules at densities up to $\sim10^{7}$~cm$^{-3}$ due to the low trappable flux and short loading times ($\sim20$~ms) attainable when loading single pulses of molecules. While the first interactions between laser-cooled molecules were recently observed \cite{Anderegg2019}, many proposed applications require larger trapped samples at higher density and increasing the trappable flux remains a key challenge. More efficient slowing techniques are currently being pursued by multiple groups \cite{DeMille2013,Fitch2016,Aldridge2016,Yang2017,Petzold2018}, and the production of brighter, slower molecular beams remains an active and complementary area of research \cite{West2017,Singh2018,Xiao2019,Jadbabaie2019,Gantner2020}.

This Rapid Communication presents a cryogenic source capable of producing bright, continuous beams of cold free radicals via laser ablation, thereby realizing the first step towards longer MOT loading times and the continuous accumulation of conservatively trapped dark-state molecules using an intermediate MOT stage. Our source uses helium buffer gas at $2.6$~K and produces 20~ms duration pulses of molecules at repetition rates up to 55~Hz, limited by our ablation laser. To the best of our knowledge, these are the brightest time-averaged beams of free radicals reported from a helium buffer gas source and mark the first demonstration of a continuous beam of cold free radicals.

The heart of our source is a two-stage pulse-tube refrigerator (Cryomech PT420) paired with a closed liquid helium reservoir \cite{Cryomech1} between the refrigerator's second-stage and the cooled copper cell (fig. 1a). When cold, the reservoir contains $\sim7$~g (1.7 moles) of helium which acts as a large thermal mass (heat capacity $\sim16$~J/K) to both dampen temperature oscillations from the refrigerator and allow the source to absorb high thermal loads from the ablation laser with limited heating. A constant source temperature is desirable to ensure reproducible molecular beam properties including flux, forward velocity and rotational state distribution. We note that an equivalent thermal mass using copper alone at 2.6~K is impractical and would require cooling $\sim400$~kg of material. However, rare-earth alloy plates have been successfully used to dampen thermal oscillations in a similar manner \cite{Allweins2008}. While beneficial once cold, the closed helium reservoir leads to longer source cool-down and warm-up times. To counter this increase, our design limits the additional thermal mass to 6~kg of machined aluminum and copper parts while largely replicating the source geometry of ref. \cite{Barry2011} (fig. 1b). Our design cools from 295~K to 2.5~K in $\sim2$~hours and can warm-up to 280~K in $\sim4$~hours, allowing rapid prototyping (fig. 1c). To help reduce warm-up times, we typically apply $0.5~$W of 808~nm laser light to the cell to increase the liquid helium evaporation rate. At base temperature, the cell and refrigerator second-stage are stable to $\pm5$~mK and $\pm60$~mK respectively (fig. 1d). For reference, without the helium reservoir the second-stage temperature stability is typically $\pm200$~mK as the refrigerator pulses \cite{cryomech_pc}. These larger oscillations have been reported to correlate with a $\sim25~\%$ peak to peak variation in molecular beam flux \cite{Barry2011}, forcing several experiments to synchronize their repetition rates to the period of the pulse-tube refrigerator to recover reproducible pulses of molecules \cite{Barry2012,Norrgard2017}.

 \begin{figure}[t b]
\centering
\includegraphics[scale=.19]{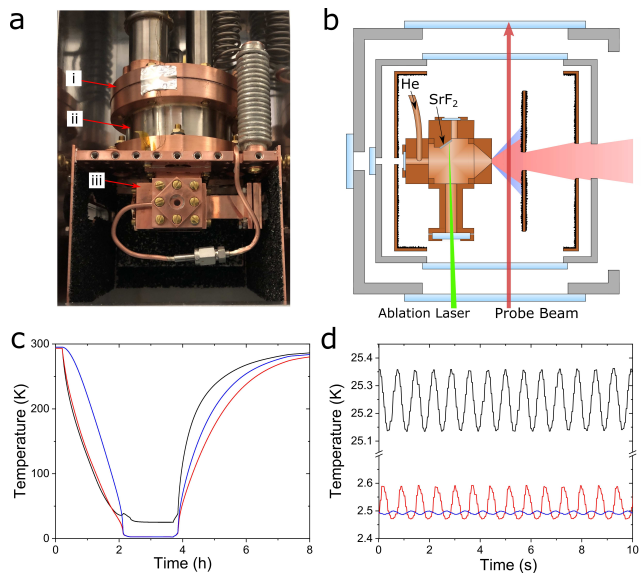}
\caption{Overview of the cryogenic source. \textbf{a} Photo of the source from behind with the rear radiation shields removed to show (i) the refrigerator second-stage, (ii) liquid helium reservoir and (iii) cell. \textbf{b} Schematic cross-section of the source from above showing the ablation and absorption beam paths. \textbf{c} Typical source cool-down and warm-up curves measured over several hours for the refrigerator first-stage (black), second-stage (red) and cell (blue). \textbf{d} Short-term temperature stability for the same three regions as \textbf{c}. Temperature oscillations at the 1.4~Hz period of the pulse-tube refrigerator are visible at all three regions.} \label{fig:Source}
\end{figure}

This work uses SrF molecules to characterize the source performance by ablating a SrF$_{2}$ target, mounted inside the cell at $30^{\circ}$ relative to the molecular beam axis, using $15$~mJ pulses of 532~nm light with $6$~ns duration from a Nd:YAG laser. This light is tightly focused onto the surface of the target using a 200~mm focal length lens outside the vacuum chamber and the pulse energy is stable to within $1~\%$. Cold helium buffer gas enters the cell through a fill line at the rear and exits through a conical face with a $40^{\circ}$ half-angle and a 3~mm diameter aperture (fig. 1b). The typical helium buffer gas flow rate is 15 standard cubic centimeters per minute (sccm), equivalent to an in-cell steady-state helium density of $10^{16}$~cm$^{-3}$ and a Reynolds number of $\approx60$. At this flow rate the vacuum inside the cryogenic source chamber is $10^{-7}$ Torr, maintained by $\sim$700~cm$^{2}$ of cold charcoal cryopump.

Properties of the molecular beam are typically probed on the X$^{2}\Sigma(v=0, N=1) $ to A$^{2}\Pi_{1/2}(v'=0, J'=1/2)$ transition at 663~nm using either absorption 20~mm downstream of the cell exit aperture or fluorescence detection 940~mm downstream. The molecular beam forward velocity is $\approx140~$m/s with a FWHM of $\approx50~$m/s. This was measured through the Doppler shift between two fluorescence profiles recorded using probe lasers transverse and counter-propagating to the molecular beam. The measured FWHM transverse velocity spread is 80~m/s, corresponding to a FWHM angular spread of $30^{\circ}$. The rotational temperature of the molecular beam is $1.0(2)$~K, measured by extracting the relative populations in X$^{2}\Sigma(v=0, N=0-4)$ using fluorescence signals from the X$^{2}\Sigma(v=0, N=0-4) $ to A$^{2}\Pi_{1/2}(v'=0, J'=1/2-9/2)$ transitions and calculated branching ratios to account for the varying line strengths \cite{Sauer1996}. Here molecules cool rotationally to below the cell temperature due to isentropic cooling near the cell aperture \cite{Pauly2000} and at our rotational temperature $\approx50~$\% of the molecules populate the X$^{2}\Sigma(v=0, N=1)$ state. All of these parameters are in good agreement with measurements performed on a source with similar geometry \cite{Barry2011}.

\begin{figure}[b!]
\centering
\includegraphics[scale=.33]{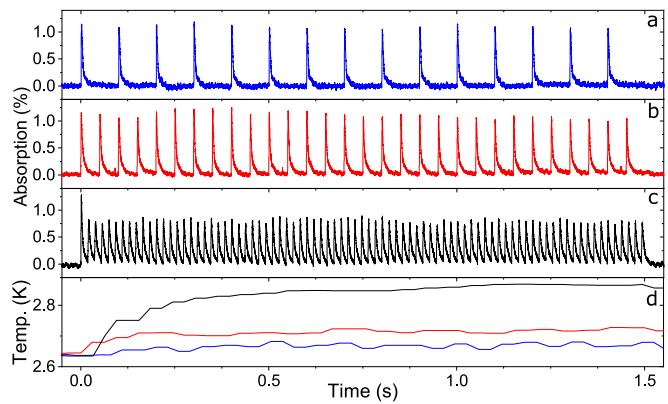}
\caption{Absorption and cell temperature traces measured over 1.5 seconds of source operation for ablation rates of (a) 10~Hz (blue), (b) 20~Hz (red) (c) and 55~Hz (black). During these measurements the source temperature (d) increased by 40, 80 and 200~mK respectively. The time needed to cool back to 2.64~K was 1~s, 30~s and 70~s for 10, 20 and 55~Hz operation respectively.} \label{fig:Abs1}
\end{figure}

To extract the number of molecules exiting the source in the X$^{2}\Sigma(v=0, N=1)$ state we use the time-integral of the resonant absorption signal, Doppler broadened absorption cross section \cite{Budker2008}, and assume a uniform density over the cross sectional area of the molecular beam \cite{Barry2011}. At ablation repetition rates of 1-2~Hz, where other helium buffer gas sources typically operate \cite{Barry2011,Iwata2017,Truppe2017}, the source produces $10^{11}$ molecules per steradian per pulse with negligible heating. We note that all reported numbers can vary by $\approx\pm50~$\% depending on the spot ablated on the target. At ablation repetition rates of 10 and 20~Hz the pulses of molecules are unchanged and the source produces $10^{12}$ and $2\times10^{12}$~time-averaged molecules/sr/sec respectively (figs. 2a and 2b). At 55~Hz we consistently observe a $\sim10~\%$ decrease in brightness ($\propto$ the time-integrated absorption signal) over the first 2-5 pulses, with negligible change in rotational temperature, and typically produce $5\times10^{12}$~molecules/sr/sec (fig. 2c). In-cell absorption measurements show that this initial decrease in brightness is correlated with decreasing in-cell molecular density and the extraction efficiency from the cell remains unchanged at $\sim50$~\%. This decrease in yield is temporary and a $100~$ms pause in ablation pulses is sufficient to recover the original yield from the next pulse using the same ablation spot. This behavior is presumably due to heating within the cell, which is measured to increase by 40, 80 and 200~mK over 1.5~seconds of operation at 10, 20 and 55~Hz respectively (fig. 2d). Once the cell reaches thermal equilibrium, we typically measure a temperature increase of $\approx1~$mK/mW of incident power, corresponding to a steady-state cell temperature of 3.5~K at 55~Hz.

\begin{figure}[b!]
\centering
\includegraphics[scale=.43]{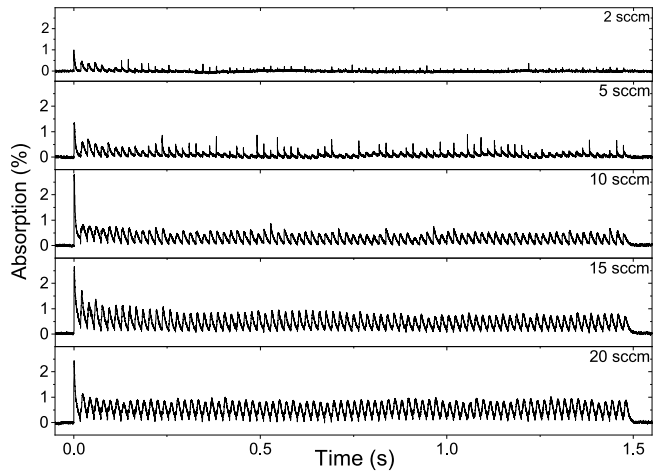}
\caption{Absorption versus helium buffer gas flow at a 55~Hz ablation repetition rate. From top to bottom, the helium flow rates were 2, 5, 10, 15 and 20~sccm. At 2 and 5~sccm some ablation pulses produce $\sim100~\mu s$ bursts of molecules, these are not detected at higher flow rates. Molecules are continuously detected leaving the source for flow rates $\geq10~$sccm. These data were recorded in a random order using the same ablation spot on the target and highlight the temporary nature of the decrease in brightness measured over the first 2-5 pulses.} \label{fig:HeFlow}
\end{figure}

The successful production of bright, continuous beams of cold free radicals via ablation at 55~Hz depends critically on the helium buffer gas flow rate through the cell (fig. 3). At flow rates of 2 and 5~sccm, the initial yield per ablation pulse is reduced by factors of 10 and 5 respectively relative to our standard 15~sccm flow rate. We attribute this decrease to insufficient buffer gas density for complete thermalization. At 2~sccm, few molecules are detected in the beam after $250~$ms of operation while a flow of 5~sccm is sufficient to consistently produce pulses of molecules and realize $10^{12}$ molecules/sr/sec. Note that the sporadic spikes in fig. 3 visible at flows of 2 and 5~sccm are $\sim100~\mu s$ pulses of $10^{8}-10^{9}$ molecules and are not optical pickup due to the ablation laser. At flows of 10~sccm we begin to continuously detect molecules exiting the cell with a brightness of $3\times10^{12}$ molecules/sr/sec, increasing to $7\times10^{12}$ molecules/sr/sec at 20~sccm. At 15~sccm the beam brightness is typically modulated at 55~Hz by $\sim80~\%$ 20~mm downstream of the cell. We project that this modulation will decrease to $\sim60~\%$ a further 1.5~m downstream by convolving the measured molecular pulse temporal and velocity distributions.

The source performance is robust at 55~Hz for helium buffer gas flow rates $\geq10~$sccm while a flow rate of $\geq5~$sccm is sufficient to sustain continuous operation at 10 and 20~Hz. Other groups using helium-based cryogenic sources typically use lower buffer gas flow rates, between 1-5~sccm  \cite{Barry2011,Truppe2017,Anderegg2017,Collopy2018}, and several have reported erratic source behavior for ablation repetition rates $>5~$Hz  \cite{Barry2011,Truppe2017}. We assume these observations are due to insufficient helium flow and, other than the ablation pulse energy, we have found no other parameter that strongly effects beam brightness. Our source performance does not depend critically on the cell geometry; molecular beams with similar brightness were realized when the conical cell exit was replaced with a flat 0.5~mm thick copper plate containing a 3~mm diameter aperture. This lack of dependence on cell geometry is in contrast to observations for a capillary fed cryogenic source \cite{Singh2018}. Similar source performance was also realized for an elevated cell temperature of $4$~K, after accounting for an increased rotational temperature of $2.0(3)~$K, which reduced the number of molecules in the X$^{2}\Sigma(v=0, N=1)$ state by $\approx30~$\%. This factor of 2 increase in rotational temperature highlights the importance of limiting heating during source operation to ensure reproducible beam properties. This source design has also proven to be straightforward to replicate and a second unit is now operational in our group and produces similar continuous beams of molecules.

As a demonstration of the stable and continuous nature of these molecular beams we produce uninterrupted pulses of SrF molecules at 55~Hz over a 60~second duration using a buffer gas flow rate of 15~sccm  (fig. 4). During this time the ablation spot was moved every $\sim10~$s to restore the decaying ablation yield and we measure a mean brightness of $\sim3\times10^{12}$ molecules/sr/sec alongside a  cell temperature increase of 0.6~K. Currently the main limit on beam brightness during continuous operation of our source is the durability of the ablated target. Free radical production methods that ablate metals in the presence a reactant gas (e.g. SF$_{6}$) have been shown to produce brighter, more reproducible beams \cite{Truppe2017}, and could potentially work well with our source design at high ablation repetition rates. Assuming a durable target is used, the absolute limit on operation time is set by saturation of the charcoal cryopump. In our current design, a continuous  flow of 15~sccm of helium can be maintained for $10~$hours before saturation and this is readily extended by increasing the cryopump surface area.

\begin{figure}[t!]
\centering
\includegraphics[scale=.43]{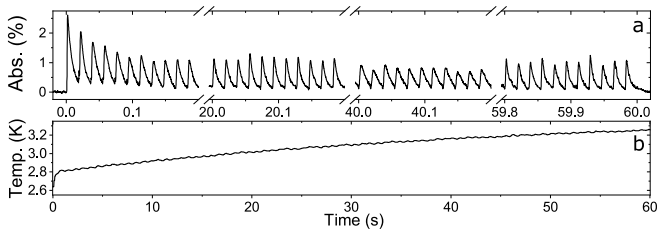}
\caption{A continuous beam of SrF molecules produced over 60~s showing (a) absorption and (b) the cell temperature. The mean brightness measured during this minute was $\sim3\times10^{12}$ molecules/sr/sec and afterwards the cell required $\sim2~$ minutes to cool back down to 2.6~K.} \label{fig:10sec}
\end{figure}

In general, beams of molecules from helium buffer gas sources are preferable over neon-based beams for molecular laser cooling and trapping experiments due to their lower forward velocities, reduced divergence and colder rotational temperatures \cite{Patterson09,Hutzler2011}. Our source design combines these advantages alongside the ability to absorb high thermal loads with limited heating, similar to neon-based sources. We note that while the large thermal mass in our design restricts heating, the maximum input power is determined by the cooling power of the pulse-tube refrigerator. In our setup the maximum load is 2~W, permitting ablation repetition rates beyond 100~Hz and possible access to beams with more than $10^{13}$~molecules/sr/sec, similar to the brightest beams of free radicals from neon-based sources \cite{Hutzler2011b}. At present, liquid helium fills $\sim25~\%$ of the closed reservoir and increasing the helium mass would further improve the source temperature stability at the expense of longer cool-down and warm-up times. In principle, one could also pump on this reservoir to cool the source towards 1~K and produce colder, slower molecular beams, but with substantially reduced cooling power \cite{Patterson09,Wang2014}.

In summary, we have realized a robust cryogenic buffer gas source capable of producing continuous beams of cold free radicals via laser ablation with a time-averaged brightness of up to $7\times10^{12}$ molecules/sr/sec in a single rovibrational state. Crucially, the performance of our source does not depend critically on the cell geometry or temperature. This suggests that other groups currently using helium-based sources could immediately adopt our method to realize brighter molecular beams, provided that there is sufficient buffer gas flow and operation times are short to limit heating. These molecular beams represent the first step towards longer MOT loading times to trap larger samples at higher density and are well-suited for continuous beam slowing and cooling techniques such as centrifugal deceleration \cite{Chervenkov2014,Wu2017}, Zeeman slowing \cite{Petzold2018}, and Zeeman–Sisyphus deceleration \cite{Fitch2016}. For reference, today's molecular MOTs load single pulses of molecules over $\sim20$~ms. If similar loading times were used for atomic MOTs only $\sim10^{5}-10^{6}$ atoms would be trapped \cite{Xu03,Harris2008}. Given that typical molecular MOT lifetimes are short, $\sim100~$ms \cite{McCarron2018}, the continuous accumulation of conservatively trapped molecules via an intermediate MOT stage is a particularly promising approach, a method which has been successfully demonstrated with chromium atoms \cite{Schmidt2003}. These advances have the potential to increase the numbers and densities realized in laser-cooled samples of molecules by orders-of-magnitude and provide routine access to the molecule-molecule interactions required for many proposed applications.

Drawings of the machined parts needed to replicate this cryogenic source design and assembly instructions can be provided by contacting the corresponding author.
\\

We thank D. DeMille and M. Steinecker for helpful discussions and J. Schnaubelt for carefully reading the manuscript. We acknowledge financial support from the NSF (CAREER Award \#1848435) and the University of Connecticut, including a Research Excellence Award from the Office of the Vice President for Research.

\bibliography{thebib.bib}

\end{document}